# Search Strategies of Library Search Experts

**Kristiina Singer, Georg Singer, Krista Lepik, Ulrich Norbisrath, and Pille Pruulmann-Vengerfeldt**

University of Tartu

**Abstract:** Search engines like Google, Yahoo or Bing are an excellent support for finding documents, but this strength also imposes a limitation. As they are optimized for document retrieval tasks, they perform less well when it comes to more complex search needs. Complex search tasks are usually described as open-ended, abstract and poorly defined information needs with a multifaceted character. In this paper we will present the results of an experiment carried out with information professionals from libraries and museums in the course of a search contest. The aim of the experiment was to analyze the search strategies of experienced information workers trying to tackle search tasks of varying complexity and get qualitative results on the impact of time pressure on such an experiment.

**Keywords:** Information professionals, search strategies, library search contest

## 1. Introduction

Despite the fact that nowadays a certain amount of duties of library professionals can be automated by using modern search technologies, such human search experts are still the key intermediates between information seekers and the information repositories in a modern library environment. Dent (2007) summarized the main flaws of search technologies the following way: Their implementations are neither able to speak, nor able to comprehend content in any way. They are not able to draw connections between sources and not able evaluating the quality of a source.

In our times of information overload, users need the assistance of library search professionals more than ever to help them find high quality resources (Han & Goulding 2002). Although it might seem questionable in the light of theories on information society (Webster 2006), the common denominator for people who work in libraries, museums or archives is being an *information professional*. These persons "have to deliver information, information products and information services for special problem situations in which users seek for information. Information professionals are supposed to navigate users to resources. Information professionals should also help determine information needs of users" (Steinerova 2001). Information professionals in the context of our work are characterized not only by their ability to search information, but also by other skills of information literacy like being able to represent the problem space and possible solutions.

Those abilities make this profession a very interesting research target when it comes to learning more about search strategies. As Kuhlthau (2005) has pointed out in his Information Search Process framework, the process of information seeking is "based on four criteria: task, time, interest, and availability" and "one of these may predominate at any given time". In the context of our study where the task is pre-given, lack of time can be an important factor that can have an impact on the performance of solving information search tasks.

In this paper, we will present the results of a search experiment carried out with information professionals from libraries and museums in the course of a search contest. We will present a new taxonomy of search strategies and apply it to our experiments. We will show what strategies those experienced information workers chose in order to solve the complex problems assigned to them during the search contest and that those information professionals are preferably applying one type of search strategy. The paper will also show that a certain Internet user type performs significantly better than the others.

## 2. Related Work

Search strategies have increasingly been researched in the last years. Marchionini (1995) has defined four levels of description in information seeking: moves, tactics, strategies, and patterns. He defined *strategies* as generalized approaches to particular information seeking problems. As proper studies about search strategies of information professionals rarely exist, we focus on the work that was done on search strategies in general. Navarro-Prieto et al. (1999) identified bottom-up, top-down, and mixed strategies. A *top-down strategy* means that users search in a general area and then narrow down their search from the links provided until they find what they are looking for. In the *bottom-up strategy* users look for the specific keyword that was provided to them in the instructions. This strategy was most often used by experienced participants, for the specific fact-finding searches. Chin & Fu (2010) found in their study that younger users prefer the bottom-up interface-driven strategy. They look up more links and leave a web page quickly. Older users prefer the top-down knowledge-driven strategy. They look up only a subset of links, take longer time to click a link, and leave a web page later.

Thatcher (2008) has studied cognitive search strategies among experienced and less experienced web users. He identified the following cognitive search strategies: (1) parallel player, (2) parallel hub-and-spoke, (3) known address search domain, (4) known address, (5) virtual tourist, (6) link-dependent, (7) to-the-point, (8) sequential player, (9) search engine narrowing and (10) broad first. Participants with higher levels of web experience were more likely to use strategies 1-4, whereas participants with lower levels of web experience were more likely to use strategies 5-10. This system contains numerous overlaps among the strategies mentioned and can therefore be used more for characterization than for a classification. We therefore created a classification of search strategies (better meeting our requirements), which we present in the next section.

The concept of a *search task* is important especially when it comes to search strategies. Schneiderman (1997) distinguished searching tasks from specific fact-finding to more unstructured open-ended general-purpose browsing tasks. The latter are usually classified as *exploratory search tasks*. Those were investigated extensively by Marchionini (2006) and White & Roth (2009) and are usually described as open ended, abstract, and poorly defined information needs with a multifaceted character.

*The Search-Logger framework* (Singer at al. 2011) that we used in our experiment is a research framework to monitor and analyze search tasks. It consists of a plug-in for Firefox web browsers and a corresponding database back-end. The Search-Logger collects implicit user information by logging a number of significant user events like links clicked, queries entered, tabs opened and closed, bookmarks added and deleted, and clipboard events. Explicit information is gathered via user feedback in the form of customizable questionnaires before and after each search task. We selected Search-Logger as the framework to monitor the contest over other web monitoring tools as Search-Logger is the only one, which supports exploratory search tasks. Furthermore, it is developed by our research group and therefore was easy to adapt to our monitoring needs.

We are also interested in the correlation between the specific Internet user types and our experiment results. Kalmus et al. (2009) define in their work the following types of Internet users: *Active versatile* (these are more active using different Internet possibilities like communication, information and entertainment compared to other groups), *entertainment-oriented active* (mainly on searching for entertainment, and consumption of culture), *practical work-related* (focus on information and practical activities, active in using e-services), *practical information-oriented small-scale* (slightly higher than average use of information and e-services), *entertainment and communication-oriented small-scale* (searching for entertainment, communication, passive Internet use with regard to other purposes) and *small-scale* (not characterized by any specific Internet use, very lowly developed online behavior).

## 3. Methodology

We conducted our study within the framework of an information search contest that is carried out annually among library and museum search experts in Estonia. The contest consisted of two rounds: from 50 participants in the first open round, the ten best in terms of accuracy of their solutions (but only one representative per institution) were selected for the second round. We used the second round of the contest to carry out our experiment. It took place in a laboratory environment in class, where Search-Logger (as outlined in the related work section) was pre-installed and pre-activated on ten computers. The demographics of our user sample is summarized in Table 1. It consisted of 10 women, aged between 27 and 51.

The search contest consisted of 15 search tasks (the answer had to be available somewhere in public websites) that can be classified as exploratory search tasks of varying complexity as defined in the related work section, and lasted for two hours. The complete list of questions can be obtained from the Search-Logger web page (Singer and Norbisrath (2010)) upon registration. The following five questions are some examples of the tasks assigned during the contest:

- (2) Find open access journals (that need not to be scientific journals) that are dedicated to school librarians.

- (3) Who is in the picture and which Austrian writer for children is the author of this little fellow? (Illustration omitted due to copyright reasons)

- (5) How to calculate the area of this figure? Please find an appropriate formula!

- (8) What kind of natural paint could you use for painting both wooden furniture and walls of the room?

- (9) Let's suppose that the building of a small village library is quite in a bad condition and desperately needs some renovation. Who and from what program could apply for funds for renovation works?

For our main analysis on search strategies we created a new classification of search strategies (see Table 2) which is based (1) on the work by Thatcher (2009) described in our related work section and (2) on the search behavior of the information professionals taking part in our experiment. The new classification consists of two main groups: "*Known address*" strategies and "*Search engine*" strategies as people generally start their search with one of these options. Known address strategies are characterized by users directly navigating to a web site (not a search engine) they already know about.

*The Known address* group comprises the following three subgroups: "*search terms narrowing*" (users carried out the search using the search function on the known web page), *"narrowing in categories"* (users are clicking through the category structure or directory structure of a web site) and *"to the point"* (users first used the specific search terms to get directly to the answer).

*The Search engine* group comprises the following four subgroups: *"search terms narrowing"* (users carrying out the search using search engine results only), *"search terms narrowing and extending"* (users navigating to a search engine, starting with a specific search term and then using a certain result to broaden the search with a query based on this result), *"search terms narrowing in categories"* (users using a special category of a search engine (e.g. images, news or products) and carrying out the search in this limited domain) and *"narrowing in categories"* (users using a special category or directory of the search engine and narrowing down this category without using the query function of a search engine).

In addition, we have also analyzed the efficiency of the whole search process, checking if the participants only worked with one browser tab or used multiple browser tabs ("*parallel-player strategy"* or not).

## 4. Results

As stated in the last section we have analyzed the recorded data to identify the search strategies (as outlined in Table 2) of library and museum professionals. First, we divided the users according to the Internet user types we outlined in the previous section. In Table 1 we have listed the users and their respective Internet user types of the participants in the sequence of how they scored in the contest. This way we can observe correlations between Internet user type and their score in the contest. It is interesting to see that the participants who scored first, second, and third in the contest all have an Internet user profile "active versatile". This Internet user profile is the most active one, having high scores on all dimensions (as outlined in the related work section). We can conclude from this that the more users are active in different areas of the Internet the more they have Internet usage experience and the better (quicker, more efficient) they are in information seeking and problem solving on the Internet.

Table 1 also shows how many points the participants scored in the experiment, the rank they achieved in the contest, and age and gender of the contest participants. The last two columns summarize the basic search performance related measures like the average number of tabs opened and closed and the average number of words used per user for all queries in the experiment. The last row of Table 1 shows averages of all measures over all users and all search tasks.

As we can see from Table 1 the winner used the least amount of 35 tabs over all and typed in the longest average search queries. It appears that the winner used a very efficient strategy, applying long and very relevant queries. User 8 on the second place used the second longest average query length after the winner.

From this we can conclude that longer and very relevant queries can be seen as an indicator for being successful and efficient in information seeking on the Internet, at least in exploratory search tasks. An average of 16 points out of 30 per user means that the contest questions were obviously challenging to answer for the contest participants.

| User | Score | Rank | Age | Gender | Internet user type | Number of opened and closed tabs | Average number of words in search string |
|---|---|---|---|---|---|---|---|
| User 3 | 28 | 1 | 40y | f | active versatile | 35 | 2.2 |
| User 8 | 23 | 2 | 25y | f | active versatile | 43 | 1.9 |
| User 2 | 21 | 3 | 36y | f | active versatile | 70 | 0.9 |
| User 9 | 16 | 4 | 35y | f | practical work-related | 55 | 1.7 |
| User 5 | 15 | 5 | 27y | f | active versatile | 61 | 1.7 |
| User 1 | 15 | 6 | 36y | f | practical information-oriented small-scale | 39 | 1.5 |
| User 10 | 14 | 7 | 48y | f | practical work-related | 45 | 1.8 |
| User 7 | 13 | 8 | 27y | f | active versatile | 59 | 1.7 |
| User 6 | 10 | 9 | 51y | f | practical information-oriented small-scale | 35 | 1.7 |
| User 4 | 9 | 10 | 49y | f | practical information-oriented small-scale | n.a. | n.a. |
| **Average** | **16** | | | | | **49** | **1.7** |

Table 1 Internet user types (n.a. means that the data was not available)

In the next step we have analyzed the data regarding the search strategies that were applied by the contest participants. Table 2 contains the absolute counts for how often each strategy was used by each user throughout the experiment. The last column states the total number of tries for the whole experiment. It can be seen that the participants who scored first and second used quite different approaches. While the winner, User 3, only needed 57 tries, the second, user 8, needed 167 tries. Obviously User 8 much more applied a trial-and-error like approach, while the winner was very efficient with fewer and more precise queries. It could possibly also be explained by the age of the users as some previous research showed (see Chin et al. in related work section) that younger users look up more links and leave a web page more quickly than the older users. The last row of Table 2 "Distribution" illustrates how often a certain type of strategy was applied in relation to the total number of trials.

Overall the library search experts most often applied a search engines strategy with subtype "search terms narrowing" (84,6%). This was followed by a search engine strategy with subtype "search terms narrowing extending" (7,4%). Only on the third place was the first "known address" strategy with subtype "search terms narrowing" (3,1%). It seemed that if users were aware of a web site, where they expected to find the information, a "known address" strategy was tried first and when it failed, a search engine strategy was used as a back-up.

The predominant use of a search engine strategy (with subtype "search terms narrowing") means that they start with a search term and narrow down the result space until they have found the information needed. Also the second most applied sub strategy "search terms narrowing-extending" is related. Here the users are alternatively getting narrower and broader again by using a certain search result as the basis for formulating new queries in order to explore a bigger result space. Overall all the contestants used a parallel-player strategy throughout the whole experiment, continuously having multiple browser tabs open and closing old ones and opening new ones which is common for more complex exploratory search tasks. Our experiment has shown that exploratory search tasks are too complex to successfully be solved by one single strategy. Which one is used depends on the task, on the skill set and knowledge of the person.

|  | Search engine | | | | Known address | | | | |
| --- | --- | --- | --- | --- | --- | --- | --- | --- | --- |
|  | search term(s) -> narrowing | search term(s) -> narrowing-extending | search term(s) -> narrowing in categories | narrowing in categories | search term(s) narrowing | narrowing in categories | to the point | Total number of tries | Parallel Player Strategy |
| User 3 | 39 | 10 | 3 | 0 | 2 | 2 | 1 | 57 | yes |
| User 8 | 151 | 5 | 2 | 0 | 3 | 4 | 0 | 165 | yes |
| User 2 | 82 | 6 | 0 | 0 | 1 | 0 | 0 | 89 | yes |
| User 9 | 115 | 9 | 1 | 2 | 1 | 0 | 0 | 128 | yes |
| User 5 | 73 | 12 | 0 | 0 | 3 | 8 | 0 | 96 | yes |
| User 1 | 32 | 3 | 2 | 3 | 3 | 1 | 0 | 44 | yes |
| User 10 | 80 | 5 | 0 | 1 | 4 | 1 | 0 | 91 | yes |
| User 7 | 112 | 10 | 4 | 0 | 2 | 0 | 0 | 128 | yes |
| User 6 | 64 | 4 | 2 | 0 | 10 | 5 | 0 | 85 | yes |
| User 4 | 44 | 5 | 3 | 0 | 0 | 0 | 1 | 53 | yes |
| **Total** | **792** | **69** | **17** | **6** | **29** | **21** | **2** | **936** | |
| **Distribution (%)** | **84.6** | **7.4** | **1.8** | **0.6** | **3.1** | **2.3** | **0.2** | | |

Table 2 Strategies applied by contest participants (ordered by final score)

In order to investigate the time pressure impact on the contest outcomes we compared the results of the contest with the results the contestants achieved in the pre-round. As opposed to what we initially expected when setting up the experiment, we could not prove a significant impact of the time pressure on the contestants. Qualitatively the results are mixed, some contestants did not show any effect, some performed better and some performed worse.

## 5. Conclusions and future work

We conducted a study about search strategies of library and museum search experts with 10 participants taking part in a search contest. It took place in a laboratory environment in class. We presented and analyzed selected search strategies of these library and museum professionals and related them to the actual behavior carrying out the respective search tasks. We also showed the relation between the observed behavior and their respective Internet user type classification.

All participants finished the contest, with the winner scoring 28 out of 30 points. The most important observation was that all participants predominantly (in 94.4% of the cases) used search engine strategies. In only 5.6% of the cases, the library search professionals applied a non-search engine strategy. This reconfirms that search engines are a good entry point to exploring a search space. Yet in case they are not sufficient they need to be augmented with specialized search portals. The low average number of points (16 out of 30) and the high average number of opened and closed tabs (49) and strategy tries per task (62) appear to indicate that search engines are not very well suitable to carry out exploratory search tasks of the kind used in the experiment. It also indicates, that the more complex an information seeking task becomes, the less search engines alone without a certain amount of personal experience and knowledge are enough to ensure problem solving.

The experiment had some limitations that we will try to resolve in future experiments. First, we only had a small sample of 10 users. So the results have more qualitative than quantitative character. Another limitation was that our sample only consisted of women. In the future we will try to also carry out an experiment with men to analyze the gender impact on the search strategy selection. For this study we have intentionally omitted the chronological order in which the strategies were applied. We have only looked at what strategies the study participants have used not taking into consideration when and in what sequence they were used. A follow up paper will add the time dimension to the results. The experiment was carried out under time pressure. Although we could not show a significant impact of time pressure on the study results, the results might have turned out differently under open ended conditions as in the experiment described by Singer et al (2011) where the study participants had 4 weeks to complete their task. We are planning an open ended follow up experiment with the same questions to further analyze the impact of time pressure on the study results. Overall the younger the participants were, the better they scored. We will also further analyze the correlation between age and search performance.

## References

Chin, J. and Fu, W-T., (2010). Interactive Effects of Age and Interface Differences on Search Strategies and Performance. In Proc. *CHI2010*, ACM Press, 403-412.

Dent, V. F., (2007). Intelligent agent concepts in the modern library. *Library Hi Tech*, 25(1),108-125.

Han, L. and Goulding, A., (2002). Information and reference services in the digital library. *Information Services and Use*, 23(4), 251–262.

Kalmus, V., Keller, M. and Pruulmann-Vengerfeldt, P., (2009). Quality of life in a consumer and information society. Lauristin, Marju (Toim.). *Estonian Human Development Report 2008*: 102 - 124. Tallinn: Eesti Ekspressi Kirjastus.

Kuhlthau, C. C., (2005). Kuhlthau's Information Search Process. Fisher, Karen E., Erdelez, S. and L. McKechnie (ed.). *Theories of Information Behavior*: 230-234. Medford, New Jersey: Information Today.

Marchionini, G., (1995). *Information seeking in electronic environments*. New York: Cambridge University Press.

Marchionini, G., (2006). Exploratory search: from finding to understanding. *Communications of the ACM*, Volume 49(4): 41–46.

Navarro-Prieto, R., Scaife, M. and Rogers, Y., (1999). *Cognitive strategies in web Searching*, http://zing.ncsl.nist.gov/hfweb/proceedings/navarro-prieto/index.html

Shneiderman, B., (1997) Designing information-abundant Web sites: issues and recommendations. In S. Buckingham Shum and C. McKnight, Eds. "Web Usability" (special issue) *International Journal of Human-Computer Studies*, 46.

Singer, G. and Norbisrath, U., (2010). *Search-Logger Website*, http://www.search-logger.com/tiki-read_article.php?articleId=2

Singer, G., Norbisrath, U., Vainikko, E., Kikkas, H., Lewandowski, D., (2011), Search-Logger - Analyzing Exploratory Search Tasks. In Proceedings of the 25th ACM Symposium On Applied Computing, TaiChung, Taiwan, March 2011. ACM Press, to appear.

Steinerova, J., (2001). Human issues of library and information work. *Information Research*, 6(2), http://InformationR.net/ir/6-2/paper95.html

Thatcher, A., (2008). Web search strategies: The influence of Web experience and task type. *Information Processing and Management* 44, 1308-1329.

Webster, F., (2006). *Theories of the information society*. 3rd ed. London; New York: Routledge.

White, R. W. and Roth, R. A.., (2009). Exploratory search: Beyond the Query-Response paradigm. *Synthesis Lectures on Information Concepts, Retrieval, and Services*, 1(1):1–98.